\documentclass[prl,twocolumn,showpacs]{revtex4}
\usepackage{graphicx}
\usepackage{amsfonts}

\usepackage{amsmath, amsthm, amssymb}
\usepackage{dsfont}
\usepackage{times}
\newcommand{\ve}[1]{\mathbf{#1}}

\newcommand{\RR}{\ve{R}}
\newcommand{\rr}{\ve{r}}

\newcommand{\e}[1]{\mathrm{e}^{#1}}

\newcommand{\ybal}{$\beta$-YbAlB$_4\,$}

\begin{document}
\title{Layered Kondo lattice model for quantum critical $\boldsymbol{\beta}$-YbAlB$_4$}
\author{Andriy H. Nevidomskyy}
\email{nevidomskyy@cantab.net}
\author{P. Coleman}
\affiliation{Department of Physics and Astronomy,
Rutgers University, Piscataway, N.J. 08854, USA}

\date{\today}
\begin{abstract}
We analyze the magnetic and electronic 
properties of the quantum critical 
heavy fermion superconductor $\beta$-YbAlB$_4$, calculating 
the Fermi surface and the angular dependence of the
extremal orbits relevant to the de Haas--van Alphen measurements.
Using a combination of 
the realistic materials modeling and single-ion
crystal field analysis, we are led to propose a layered Kondo lattice
model for this system, in which two dimensional boron layers
are Kondo coupled via interlayer Yb moments in a $J_{z}=\pm 5/2$
state. This model fits the measured single ion magnetic
susceptibility and predicts
a substantial change in the electronic anisotropy
as the system is pressure-tuned through the quantum critical point. 
  \end{abstract}

\maketitle

Heavy electron materials have 
played a long-standing role in  strongly correlated electron physics,
providing key insights into emergent quantum mechanical behavior
that many hope can be scaled up in energy and temperature in
actinide and transition metal compounds. One of the remarkable aspects
of heavy fermion materials is that they can be tuned, by applying
pressure or magnetic field at low temperatures, through a quantum
critical point~\cite{gegenwart,review-lohneysen}, where they exhibit
non-Fermi liquid behavior~\cite{coleman-dyingFL, gegenwart, review-lohneysen,
  review-NFL} and a marked predisposition towards  
superconductivity~\cite{mathur,monthoux07}.
The recent discovery of a
layered Yb based heavy fermion material \ybal~\cite{nakatsuji08}
which is both
stoichiometrically quantum critical and superconducting at 
$T_{c}=80$~mK has attracted
a great deal of interest.  There are 
many examples of  heavy electron superconductivity, often 
in close proximity to a quantum critical point~\cite{gegenwart}. However, 
although heavy electron behavior and quantum critical behavior have
been observed in Ce, U and Yb materials~\cite{CeCu2Si2,mathur,
CeCu2Ge2, CeCoIn5, UBe13, UPt3, URu2Si2}, \ybal 
is the first ytterbium based heavy fermion  superconductor. 
This material raises many fascinating questions.  
Why, for example, is this system intrinsically quantum critical, with
a specific heat coefficient 
that is finite in a magnetic field, but which 
diverges as the field is removed? 

  In this paper, we present an analysis of the
magnetic and electronic properties of $\beta$-YbAlB$_4$.  We use a
combination of electronic structure calculations and crystal field
analysis to develop a simple model for the low energy physics.
Our results are consistent with the almost localized nature
of the Yb f-electrons.  In particular, we find that the electron
effective mass, as measured by the low temperature specific heat
coefficient $\gamma=C_v/T$, exceeds the band-structure value by more
than a factor of 30.  Moreover, in an LDA+U band-structure calculation, the
system is found to develop an antiferromagnetically ordered
ground-state, underlining the close vicinity to magnetic instability.

\begin{figure}[!b]
\vspace{-5mm}
\begin{center}
{\includegraphics[width=0.35\textwidth]{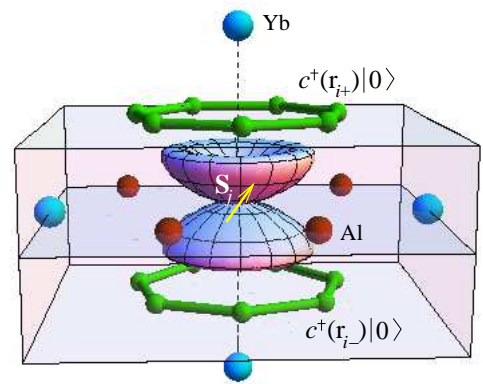}}
\end{center}
\caption{(Color online) Local coordination of Yb atom (large
  spheres, blue), suspended between two planes of boron heptagons
  (small green spheres) and surrounded in-plane by a distorted
  rectangular of Al atoms (red spheres).
 The proposed $|m_J=\pm5/2\rangle$ ground state wavefunction is also
  shown, see Eq.~(\ref{Y5/2}). Boron conduction electrons
  $c^\dagger(\rr_{i \pm}) |0\rangle$ interact with Yb spins $\ve{S}_j$
  via layered Kondo lattice model Eq.~(\ref{model}).
}
\label{Fig.struct}
\end{figure}

We have calculated the Fermi surface and the angular dependence of the
extremal orbits relevant to the de Haas--van Alphen measurements.  Despite
the layered crystal structure, hybridization with the $f$-electrons
gives rise to a three-dimensional Fermi surface.  
Combining the band-structure calculations and crystal
field analysis, we are led 
to propose an anisotropic Kondo lattice description of
this material, in which Yb magnetic moments at a site of seven-fold symmetry
hybridize with neighboring boron planes, 
to give rise to the observed 3-dimensional band structure.
We show that the crystal field configuration 
with maximum interlayer overlap, $\vert J\!=\!7/2$, $m_{J}\! =\!\pm 5/2\rangle $ 
(Fig.~\ref{Fig.struct}), accounts well for the anisotropy in the 
high temperature magnetic susceptibility.  

 The compound $\beta$-YbAlB$_4$ crystallizes in an orthorhombic structure (group
 \textit{Cmmm})~\cite{YbAlB4-struct}, with 
 ytterbium and aluminum  atoms sandwiched between boron layers.
In this unusual structure 
the
 Yb atoms are 
 centered between seven-member boron rings, as 
illustrated in Fig.~\ref{Fig.struct}, while the Al atoms are 
 centered between pentagonal rings of boron. 
A central question raised by the structure, is whether 
the electronic structure will be quasi-two dimensional as 
in the case of intercalated graphite compounds. 
Our calculations indicate instead that  the two-dimensional layers of
boron are  ``short-circuited'' by electrons of Yb and Al. The role of the Kondo
 effect in this setting is particularly intriguing and will be
 discussed below.

\begin{figure}[!tbp]
\begin{center}
\vspace{-3mm}
{\includegraphics[width=0.48\textwidth]{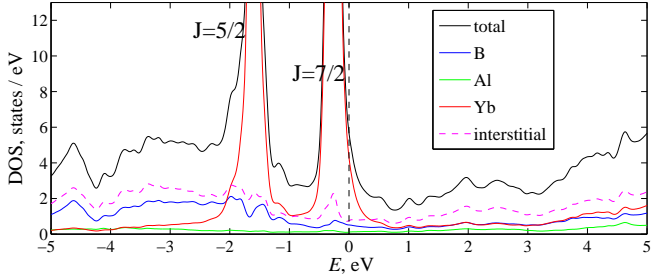}}
\end{center}

\caption{(Color online) Projected density of states (DOS), showing relative contribution of
  different atoms. 
  The DOS at the Fermi level corresponds to the specific heat coefficient
  $\gamma\equiv C_\text{v}/T=6.7$ mJ/(mol Yb$\times$K$^2$) per Yb
 atom. 
\vspace{-5mm}
 }
\label{Fig.bands}
\end{figure}

For the band structure calculations, a full potential linear augmented
plane wave method (LAPW)  was carried out using the WIEN2k code~\cite{wien2k},
using a Local Density Approximation (LDA) with a 
generalized gradient correction~\cite{PBE}. The
strong spin-orbit
coupling of the Yb atoms was taken into account  using a  scalar-relativistic approach~\cite{Singh-book,wien2k}.
The calculated electron density of states (DOS) of $\beta$-YbAlB$_4$ is shown
in Fig.~\ref{Fig.bands}.
The Yb $f$-bands are split by the spin-orbit (SO)
interaction into  $J=5/2$ and
$J=7/2$ multiplets separated by ~1.4 eV, and are heavily hybridized 
with the lighter conduction electrons of B, and (to a
lesser extent) Al atoms.
The DOS at the Fermi level is dominated by the Yb
$f$-bands, with a band mass around 5--6~$m_e$ in the
($ab$)-plane and about $m_{e}$ along the $c$-axis.
The specific heat coefficient $\gamma_\text{th}\!=\!6.7$mJ/(mol$\times$K$^2$),
extracted from the value of the DOS at $E_F$ contrasts with a
measured value $\gamma_\text{exp}=\lim\limits_{T\to 0}C_M/T \gtrsim
170$~mJ/(mol Yb$\times$K$^2$) at a field of 0.50
Tesla~\cite{nakatsuji08}. Experimentally, the zero temperature
specific heat coefficient is field dependent, diverging as 
$B^{-1/4}$. On the basis of this comparison, 
\begin{equation}
\frac{m^{*} (B)}{m_{\hbox{ band}}}
\approx  \frac{25} {(B-B_c)^{1/4}},
\label{NFL1}
\end{equation}
where the critical field $B_c=0$ within current experimental
resolution.
This divergent mass enhancement indicates that the 
Yb $f$-bands become localized in the zero field limit.

\begin{figure}[!tbp]
\begin{center}
\vspace{-3mm}
{\includegraphics[width=0.44\textwidth]{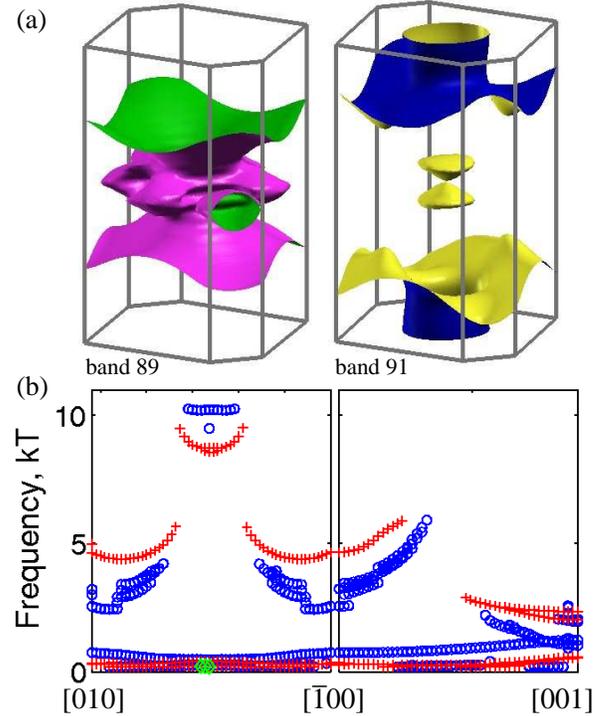}}
\end{center}
\caption{(Color online) LDA-solution calculated (a) Fermi surfaces and
  (b) angle-dependent de Haas -- van Alphen frequencies of extremal
  orbits showing only the lightest bands with frequencies
  $F\!<\!10$~kT (higher frequencies up to 50 kT are also present but are unlikely to be
  observed in dHvA measurement).
 Two Yb-$f$ bands contribute most to the Fermi surface:
  band 89 (blue circles) and 91 (red crosses).
  In addition, LDA predicts small hole pockets due to
  band 93 (green rhombi). Each band is doubly degenerate in the
  spin-orbit mixed basis, hence only odd-numbered bands are
  shown.
 \vspace{-5mm}}
\label{Fig.fermi}
\end{figure}

The LDA calculation predicts that the Yb ions are in a mixed valence
state, with $n_f\approx 13.4$. This is 
clearly inconsistent with the susceptibility measurements, which 
reveal a Curie-Weiss susceptibility at high temperatures, with the
full moment of the $4f^{13}$ state:  a signature of local moment behavior.
We attempted to cure this failure using the LDA+U
method~\cite{anisimov} where the 
Anderson $U$ is treated at the mean-field level for the localized
Yb $4f$-orbitals. 
These calculations predict a magnetic
ground state with an in-plane AFM arrangement of Yb spins and the
effective moment of 0.6~$\mu_B$ per Yb ion. 
While no  magnetic ordering is observed in
the material, we may regard the  LDA+U result as a mean-field
description of strong underlying magnetic correlations.

Although LDA  can not capture the large mass
renormalizations of strongly correlated systems, 
it generally provides a good guide to the underlying Fermi surface 
topology\cite{upt3,hitc,pnictide}. The calculated LDA Fermi surface 
$\beta$-YbAlB$_4$ (see Fig.~\ref{Fig.fermi}a))  
contains two sheets (heavy bands 89 and 91) that are $f$-like in character,
and a very small pocket (light band 93, not shown) associated with boron
$p$-electrons. The topology of the heavy sheets is non-trivial, featuring a combination of parallel quasi-1D
sheets that are connected by a cylidrical quasi-2D tube. 
For comparison with future de Haas-van Alphen (dHvA)
experiments we have determined the angular dependence of the extremal
orbits~\cite{rourke08}, shown in 
Fig.~\ref{Fig.fermi}c) for the 
the lightest bands with with LDA-calculated $m^*<10 m_e$).
The heavier bands, with larger frequencies, are unlikely 
to be seen in dHvA measurements due to
the quasi-particle damping. 

Experimentally, the low temperature
thermodynamics of \ybal is dominated by the 
spin entropy of the Yb ions\cite{nakatsuji08}, 
indicating that this system is a Kondo
lattice compound. We now 
determine the symmetry of the low lying magnetic
doublets and construct the corresponding Kondo lattice model. 
In \ybal, the 
Yb ions are suspended between two seven-member rings of boron atoms (Fig.~\ref{Fig.struct})
giving rise to an approximate 7-fold symmetry.
In this low symmetry environment, we expect the 
eight states of the $J\!=\!7/2$ multiplet
to split into four Kramers doublets. 
The approximate local symmetry group of the Yb ions is
$C_{7}\times C_{2v} $.
The only crystal field operators with both 
$2\pi/7$ rotational  and
time reversal symmetry are polynomial functions $P[J_{z}^{2}]$
of  $J_{z}^{2}$~\cite{foot-7fold}, 
so to good approximation, the magnetic doublets
are eigenstates of definite $m_{J}$.
Empirically, the observation of a Schottky peak 
in the specific heat at
$T_\Delta\approx$~30-40~K, indicates that the lowest lying crystal
field excitation has energy $\Delta\!\approx\! 2k_B
T_\Delta\!\sim$~60-80~K.

\begin{figure}[!tbp]
\begin{center}
{\includegraphics[width=0.48\textwidth]{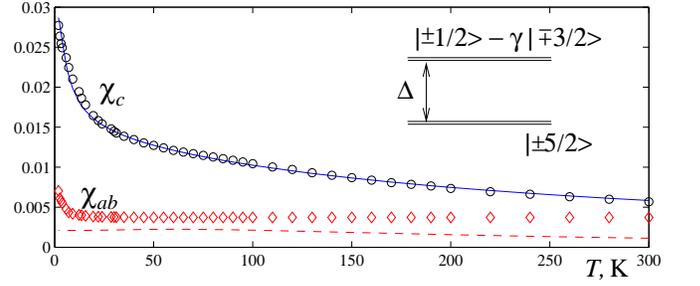}}
\end{center}
\caption{(a) Theoretical fit to experimental susceptibility
  data~\cite{nakatsuji08} $\chi_c(T)$ (circles) and $\chi_{ab}(T)$ (rhombi) in the
  range 50--300~K using the single-ion model shown in the inset and the
  fitted Curie-Weiss temperature $\theta_\text{CW}$. The
  best fit yields the lowest crystal field splitting
  $\Delta =\,$80~K, the coefficient of admixture $\gamma\approx 0.28$ and
  $\theta_\text{CW}^\text{th} = -230$~K, close to the experimental value
   $\theta_\text{CW}^\text{exp}\sim-210$~K. 
 \vspace{-3mm}
}
\label{Fig.chi}
\end{figure}

To determine the symmetry of the lowest lying Yb doublet, we 
appeal to geometric considerations. The lowest energy magnetic doublet
will correspond to the $4f$ wavefunction with 
maximum overlap with the boron rings. 
This rules out the states
with $|m_{J}|= 1/2$ which are aligned along the c-axis, and 
the states with $|m_J|=7/2$ which are planar, lying in the ab plane.  
The  remaining contenders
with $m_{J}=3/2 $ or $5/2$ are ``conical dumbbells'' aligned along the
c-axis.  The seven-fold boron rings subtend
an angle of $50^{\circ}$ about the c-axis, closely corresponding 
to the aperture angle ($52^{\circ}$) of the conical wavefunction of the 
$|\pm 5/2\rangle$ multiplet. We expect this state to have the
maximum hybridization with the boron rings.
The most likely crystal field excitation state is 
the mixture of $|\pm 1/2\rangle$ and  $|\pm 3/2\rangle$ configuration, with the second-largest overlap.
This simple model of the Yb ion provides a good fit to the 
experimentally measured magnetic susceptibility at high temperatures
(Fig.~\ref{Fig.chi}), provided one fits the Curie--Weiss 
temperature $\theta_\text{CW}^\text{th}~\sim -230$~K due to
antiferromagnetic RKKY interaction between Yb moments.
The best fit is obtained with the excited state being mainly $|\pm
1/2\rangle$, with a small admixture of the $m_{J}= \pm 3/2$ as indicated in the
inset of Fig.~\ref{Fig.chi}, with $\gamma\sim 0.28$. This is likely due
to crystal fields of Al atoms that break the seven-fold symmetry of boron rings.

Given this single-ion picture, the simplest effective model for
the low-energy degrees of freedom is the one in
which low lying $|J=7/2,m_{J}=\pm 5/2\rangle$ 
$4f$ doublet interacts with the neighboring
layers of two-dimensional
boron $p$-electrons via a Kondo exchange interaction. 
Based on our calculations, we expect 
the spins on neighboring Yb ions to interact 
via an in-plane antiferromagnetic RKKY interaction $J_{ij}$.  
This leads us to propose a ``layered
Kondo lattice model'' for \ybal, $H=H_0 + H_K + H_H$, where
\begin{eqnarray}
H_0\! &=&\! 
- \sum_{p=\pm,} \sum_{\langle m,n\rangle \in (ab)} t_{m,n}\,
                      c^\dag_{\sigma}(\rr_{m p})  c_{\sigma}(\rr_{n p})  
                      \nonumber\\
H_{K}\! &=&\! J_K \ve{S}_j\! \cdot\! \left( \sum_{j,p=\pm} c^\dag_{j\alpha
  p}\boldsymbol{\sigma}_{\alpha\beta}
c_{j\beta p} + \! \sum_{j,p\ne p'} c^\dag_{j\alpha p}\boldsymbol{\sigma}_{\alpha\beta}
c_{j\beta p'} \right)  \nonumber \\
H_{H}\! &=& \! \sum_{\langle i,j \rangle \in (ab)}\!\!\! J_{ij}
\ve{S}_i\cdot \ve{S}_j \!\! \label{model},
\end{eqnarray}
where the $c^\dag_{\sigma}(\rr_{m p})$ operators create conduction electrons
with spin $\sigma=\pm 1/2 $ in the
two dimensional boron planes at site $\rr_{m \pm}=\rr_m \pm c/2$. The 
$\ve{S}_j$ are the Yb spin operators, interacting
with boron layers via the Kondo coupling $J_K$.
The $\{c_{j\alpha p}\}$ represent boron Wannier
states in the $p^\text{th}$ layer projected onto Yb site $\RR_j$  in the 
 $|\alpha=\pm\frac{5}{2}\rangle$ state, given by
\begin{equation}
c_{j\alpha,p}=
\sum_{\sigma} \sum_{m=1}^{7}   
\mathcal{Y}^{5/2}_{\alpha\sigma}(\rr_{m p}-\RR_j)\; 
c_\sigma(\rr_{m p}),
\end{equation}
where summation is over atomic positions of the seven-member boron ring
above or below the Yb site.
Here the form-factors
$\mathcal{Y}^{5/2}_{\alpha\sigma}(\hat{\rr}) = C^J_{\alpha ,\sigma} 
  Y_{l,\alpha -\sigma}(\hat{\rr}) $,
arise due to spin-orbit coupling~\cite{coqblin},
where  $C^{7/2}_{\alpha   ,\sigma}
\equiv \langle
l\!=\!3,\alpha-\sigma  ;\sigma | J\!=\!\frac{7}{2}, \alpha  \rangle$
are the Clebsch-Gordan coefficients for the $J=7/2, l=3$ state and 
$Y_{lm}(\hat \rr )$ are spherical harmonics. 
Written explicitly
\begin{equation}
 \mathcal{Y}^{5/2}_{\alpha\sigma}(\hat{\rr})=\frac{1}{\sqrt{7}}\left( \begin{array}{cc}
\sqrt{6}\, Y_{3,2}(\hat{\rr}) &  Y_{3,3}(\hat{\rr})  \\
 Y_{3,-3}(\hat{\rr}) &  \sqrt{6}\, Y_{3,-2}(\hat{\rr})  \\
\end{array} \right).
\label{Y5/2}
\end{equation}
The angular probability distribution function of the 
\mbox{$|\pm\frac{5}{2}\rangle$} doublet is given by
$\sum_{\alpha\sigma}|\mathcal{Y}^{5/2}_{\sigma\alpha}(\hat{\rr})|^{2}$
as shown in  Fig.~\ref{Fig.struct}.

The above model (\ref{model}) provides a starting  point for future
theoretical studies of \ybal. 
Here we focus on the qualitative physics of 
the model.  At high temperatures, local moments form on the Yb
ions, which interact via RKKY interaction, whose scale is however too
small to allow magnetic ordering. The high temperature electronic
behavior is then governed mainly by the ``small'' Fermi surface formed
out of the quasi-two-dimensional boron bands, coupled weakly along the
c-axis via the interplanar aluminum atoms.  As the temperature is
lowered below the Kondo temperature $T_K\sim D\e{-D/2J_K}$
(here $D$ is the conduction electron band width),
the quenched local moments will give rise to
resonant bound-states that link  neighboring boron layers, forming 
a three-dimensional Fermi surface of heavy quasiparticles.
 
%
%
%

The
observation of field-tuned 
quantum criticality in this material indicates that the 
RKKY interactions and the Kondo temperature are of comparable
size, estimated~\cite{nakatsuji08} to be of order of
100~K, a value substantially larger than
other quantum critical systems, 
such as  YbRh$_2$Si$_2$~\cite{YbRh2Si2} ($T_{K}\sim 25$~K) or
CeCu$_{6-x}$Au$_{x}$~\cite{review-lohneysen} ($T_{K}\sim 6$~K).
The enhanced Kondo temperature suggests that this system balances a
more mixed valent Yb ion with stronger RKKY interactions.
Set against these observations, 
the tiny size of the superconducting transition temperature is quite
mysterious, for a large magnetic interaction might be expected to
induce stronger pairing~\cite{mathur}. 

A striking feature of this compound is its highly anisotropic,
low-symmetry structure. 
The layered Kondo model Eq.~(\ref{model}) is akin to a quantum dot
symmetrically coupled to two leads~\cite{glazman}, where at high temperatures or in
an applied field the $f$-electrons are localized and the system is
effectively Coulomb-blockaded. 
As the Kondo effect develops at low temperature, the interlayer channels opened up by the
Yb Kondo resonances will lead to \emph{co-tunneling} processes
described by the second term in $H_K$, 
enhancing the c-axis conductivity. 
In \ybal, neighboring boron planes are weakly
coupled by the $s$,$p$ electrons of the Al and Yb atoms, leading to a
finite c-axis conductivity even in the absence of the Kondo
coupling. We expect however that the interlayer Kondo effect will
resonantly ``short-circuit'' the boron planes at low temperatures.
In the light of the above discussion, the
conventional scenario of the quantum critical point (\ref{NFL1}) would
suggest an immediate drop in $c$-axis conductivity as the system is
tuned into the magnetic phase by applied pressure. 

The most  puzzling feature of this material
is the quantum critical behavior observed 
under ambient conditions. 
It is conceivable that the system is fine-tuned to a quantum
critical point. In this case, pressure tuning
will immediately drive \ybal into an antiferromagnetic
phase, with a mass divergence following the form Eq.~(\ref{NFL1}), 
but where $B_{c} (P)$ is now finite. In this scenario, the
fact that $B_{c}=0$ at $P=0$ would be purely accidental. 
A more intriguing possibility, is that the field-tuned properties of \ybal 
are signature of a robust \emph{quantum critical phase}. 
In this case, the critical field would be expected to remain zero 
over a finite range of pressure, while the exponent
$\alpha $ may vary with pressure, 
according to 
\begin{equation}
m^{*} (B)/m_{e}
\propto B^{-\alpha (P)},
\label{NFL2}
\end{equation}
The possibility of critical metallic phases in the Kondo lattice has
recently been put forward by Anderson~\cite{anderson}.
The existence of a paramagnet over a range of pressures between
the fully developed Fermi liquid and antiferromagnet has also been proposed
in the context of ``Kondo breakdown'' models of quantum
criticality~\cite{senthil,pepin}, however in that case the
intermediate phase is a non-critical
spin liquid phase. 

The existence of  quantum critical point at zero field 
is actually well known 
in the context of the two impurity Kondo model~\cite{varma-jones,gan}, where a competition
between the Kondo effect and valence bond formation between the local
moments leads to non-Fermi liquid behavior. Could  this phase
be stabilized  by the formation of a spin liquid in the Yb layers? 

This work was funded by NSF grant DMR 0605935. 
The authors are grateful to Zachary Fisk for bringing
Ref.~\onlinecite{YbAlB4-struct} to our attention. A.H.N. thanks
P.M.C. Rourke and S. R. Julian for kindly providing the authors with
computer code for Fermi surface analysis. 
\emph{Note added in proof:} after this Letter was submitted, a
report on de Haas--van Alphen measurements by O'Farrell {\emph
  et al.} \cite{ybal-cambridge} has become available, in good agreement with our prediction
of two heavy sheets of the Fermi surface.



\begin{thebibliography}{99}
\bibitem{foot-7fold}  While combinations of $J_{\pm}^{7}$ are invariant under
rotations, they are  odd under time-reversal and thus forbidden.
\end{thebibliography}

\end{document}